\pdfoutput=1

\documentclass[%
 reprint,
 amsmath,amssymb,
 aps,
floatfix,
]{revtex4-2}

\usepackage{graphicx}% Include figure files
\usepackage{dcolumn}% Align table columns on decimal point
\usepackage{bm}% bold math
\usepackage{siunitx}
\usepackage[version=4]{mhchem}
\usepackage{float}
\usepackage[hidelinks]{hyperref}  % use for final document
\begin{document}

\preprint{APS/123-QED}

\title{Nanoimprint strain-engineering of 2D semiconductors}% Force line breaks with \\

\author{Jannis Bensmann}
\thanks{These two authors contributed equally}
\author{Robert Schmidt}
\thanks{These two authors contributed equally}
\author{Robert Schneider}
\author{Johannes Kern}
\author{Paul Steeger}
\author{Mohammad Adnan}
\author{Steffen Michaelis de Vasconcellos}
\author{Rudolf Bratschitsch}
\email{Rudolf.Bratschitsch@uni-muenster.de}
\affiliation{%
University of Münster, Institute of Physics and Center for Nanotechnology, Wilhelm-Klemm-Str.\ 10, 48149 Münster, Germany
}%

\date{\today}% It is always \today, today,

\begin{abstract}
Mechanical strain is a powerful tool to tune the optical and opto-electronic properties of atomically thin semiconductors. Inhomogeneous strain plays an important role in exciton funneling and the activation of single-photon emitters in 2D materials. Here, we create an inhomogeneous strain profile in a 2D semiconductor on a micrometer scale by a nanoimprint process. We present a nanoimprint setup, where a mold is used to apply pressure in a controlled way to a \ce{WS2} monolayer on a heated polymer layer. After printing, the strain created in the 2D semiconductor is verified by hyperspectral optical imaging. The developed nanoimprint technique is scalable and could be transferred to commercial nanoimprint machines.
%\begin{description}


%\end{description}
\end{abstract}

\maketitle

%\tableofcontents

\label{sec:maintext}

Strained semiconductors have become a popular choice for many electronic and optoelectronic applications due to their improved electrical properties, particularly in the production of high-performance transistors \cite{Lee2005Strained}. They have also seen applications in semiconductor lasers \cite{Adams1992Semiconductor}. By applying strain to the material, its electrical properties can be altered. Strain causes the atoms of the material to be displaced, creating a strain-induced electric field that affects the charge carrier mobility \cite{Sun2010Strain}. Typically, strained semiconductor devices are made of silicon, germanium, or gallium arsenide, which are grown on top of a lattice-mismatched layer. In that way, strain is naturally created. This has enabled the production of transistors that are capable of much higher switching speeds than traditional transistors, resulting in faster and more efficient computing devices \cite{Thompson200490-nm}. 

Since the first exfoliation of graphene from graphite, a new library of two-dimensional materials has been discovered \cite{Khan2020Recent}. Strikingly, the properties of 2D materials can strongly differ from their bulk counterparts. Compared to conventional semiconductors, 2D materials are often characterized by high in-plane stiffness and low flexural rigidity, so that individual atomic sheets are intrinsically capable of sustaining larger mechanical strains. Given the high flexibility of 2D membranes, mechanical deformations are particularly suited for modulating their properties. The semiconducting transition metal dichalcogenides (TMDCs) have extraordinary optomechanical properties. Mechanical strain leads to drastic changes in the electronic band structure, affecting the optical response. Intralayer excitons exhibit a strong energy shift when uniaxial tensile strain is applied \cite{Conley2013Bandgap,Schmidt2016Reversible}. Strain also modulates exciton-phonon coupling in monolayers, leading to changes in the exciton linewidth \cite{Niehues2018Strain}. Furthermore, local strain on the nanoscale leads to the activation of single-photon emitters in 2D materials \cite{Kern2016Nanoscale,Palacios-Berraquero2017Large-scale,Branny2017Deterministic,Proscia2018Near-deterministic,Li2021Scalable}. Therefore, externally applied strain is one of the most interesting tuning knobs for tailoring the optical and electronic properties of 2D materials. 

Considering the nanoscale thickness of 2D materials, traditional tensile and bending testing methods are challenging. Therefore, several new strategies to investigate the mechanical properties of 2D materials have been developed. By transferring 2D materials onto flexible substrates, homogeneous uniaxial and even biaxial tensile strain can be applied by macroscopic bending \cite{Schmidt2016Reversible,Island2016Precise} or stretching \cite{Wang2015Strain-induced} the substrate. Typically, transparent polymer substrates are used, which allow for simultaneous applications of mechanical strain and measurements of photoluminescence (PL), optical transmission/absorption, second harmonic and Raman responses of the 2D material.

The effects of spatially inhomogeneous strain are even more interesting. Smooth strain gradients can funnel excitons toward or away from deformed areas of the 2D layer \cite{Feng2012Strain-engineered,Rosati2021Dark}. Finally, steep strain gradients may result in spatial localization of excitons leading to quantum confinement effects and the activation of single-photon sources in 2D materials \cite{Kern2016Nanoscale,Palacios-Berraquero2017Large-scale,Branny2017Deterministic}. Inhomogeneous strain can form by induced wrinkling and folding \cite{Castellanos-Gomez2013Local}. However, a deterministic creation is challenging. Recently, various strategies to induce strain in a more controlled fashion in 2D materials on a micro- and nano-scale have been developed. These include the controlled bubble formation \cite{Tedeschi2019Controlled} and in particular the transfer of monolayers on pre-patterned substrates such as nanogaps \cite{Kern2016Nanoscale}, micropillars arrays \cite{Milovanovic2019Strain}, or by wrapping a monolayer around a nanowire \cite{Dirnberger2021Quasi-1D}. Dynamic creation of local strain has been achieved by microcantilevers \cite{Harats2020Dynamics} and surface acoustic waves (SAWs) \cite{Rezk2016Acoustically-Driven}. All these methods have in common, that either the substrate needs to be prepared before the 2D material is precisely transferred or the direct manipulation of the 2D material requires complex and time-consuming techniques. Therefore, they lack either sufficient controllability and repeatability or scalability for device applications. Better suited for long-term stability are methods, where the underlying substrate is manipulated after the transfer of a 2D material. Such methods include thermomechanical nanostraining \cite{Liu2020Thermomechanical} by local application of heat to a polymer substrate or by electron irradiation \cite{Du2022Strain}. Furthermore, a mechanical deformation on the nanoscale has been employed to activate single-photon emitters by indentation of the polymer substrate through the 2D material with an atomic force microscope (AFM) tip \cite{Rosenberger2019Quantum} or by nanoimprints of pillars \cite{Lai2022Single-Photon}. 

Here, we present a nanoimprint setup for creating inhomogeneous strain profiles on a micrometer scale in 2D materials. Nanoimprint lithography (NIL) is a technique that uses mechanical deformation to create submicron structures by pressing a mold onto an imprint resist \cite{Chou1996Imprint,Chou1996Nanoimprint}. This method is flexible, scalable, and cost-effective, and has been widely used in the fabrication of photonic structures and micro- and nanoscale patterns for electronic and opto-electronic devices. 

\begin{figure}[t]
    \includegraphics{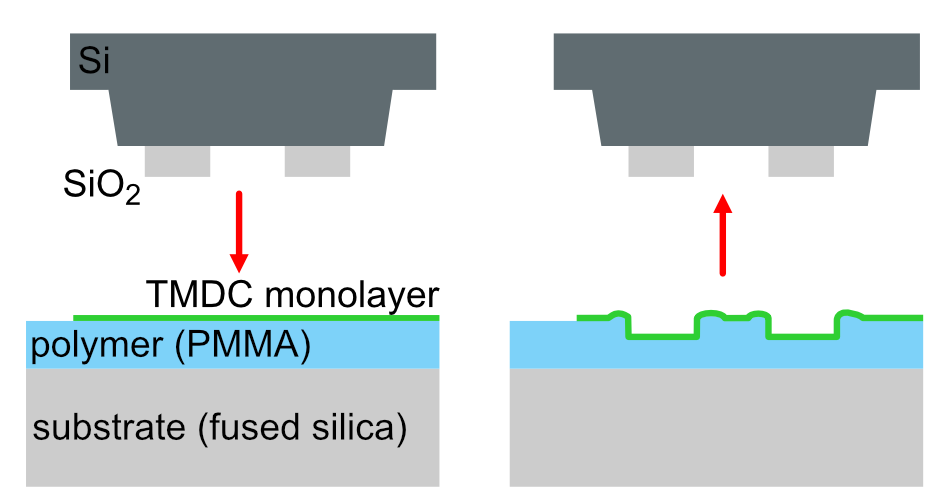}
    \caption{Nanoimprint lithography for spatially inhomogenoeus strain creation in a 2D material.}
    \label{fig:principle}
\end{figure}

To create strain profiles in mono- and multilayer TMDCs using a nanoimprint process, a sample with the 2D material on a polymer layer is prepared and the desired mold is used to apply pressure to the TMDC and polymer layer, causing them to deform and take on the shape of the mold. Since the reshaped polymer has a larger surface area than before the imprint, the monolayer, which is covering the surface, must stretch accordingly, which leads to the creation of tensile strain (Fig.~\ref{fig:principle}).

To prepare the sample, we spin-coat a fused silica substrate with a thin (\SI{270}{nm}) polymethyl methacrylate (PMMA) layer. Large-area \ce{WS2} monolayers grown by chemical-vapor deposition (CVD, Ossila Ltd., UK) are picked up using a polydimethylsiloxane (PDMS) stamp and transferred onto the thin polymer layer using dry viscoelastic stamping \cite{Castellanos-Gomez2014Deterministic}. 

The mold with the desired pattern is fabricated using electron beam (e-beam) lithography and subsequent reactive ion etching and wet chemical etching steps (see Fig. \ref{fig:process}). Initially, silicon wafer pieces with an \SI{80}{nm} \ce{SiO2} layer on top are cleaned in acetone and isopropanol in an ultrasonic bath and then heated on a hotplate to \SI{150}{\celsius} for \SI{4}{min}. In a first step, pedestals with an approximate size of \SI{100}{\micro m} and alignment markers are fabricated. To this end, an adhesion promoter (TI Prime, Microchemicals) is applied, followed by a \SI{450}{nm} thick layer of electron-beam resist (Allresist, ARN 7520.18), which is spin-coated at 4000 rpm for 60 seconds and baked at \SI{85}{\celsius} for 60 seconds using a hot plate. The resist is subsequently structured using e-beam exposure in a Raith EBPG5150 machine with a voltage of \SI{100}{kV}. The exposure is done using a current of \SI{100}{nA} and a dose of \SI{2000}{\micro C/cm^2}. All e-beam lithography patterns are designed with the python package GDSHelpers \cite{Gehring2019Python}.

After exposure, the resist is developed in MF-319 (Microposit) for \SI{60}{s} at \SI{85}{\celsius} and hardened with a postbake to increase its plasma etching resistivity. As a next step, the top \ce{SiO2} layer is etched using inductively coupled plasma - reactive ion etching (ICP RIE) in a Meyer Burger MicroSys 200.  The plasma is formed by \ce{CHF3} and \ce{O2} gas (flow rates are \SI{50}{sccm} \ce{CHF3} and \SI{5}{sccm} \ce{O2}), where fluorine ions are the reactive species and carbon deposition is minimized by reactive oxygen ions. Using a bias voltage of \SI{380}{V} and an ICP Power of \SI{100}{W} at a pressure of \SI{5e-2}{mbar}, an etch rate of \SI{34}{nm/min} is achieved. Residual resist is removed by exposing the sample to an \ce{O2} plasma (MicroSys 200, \SI{10}{min}). The area next to the pedestal is now free of resist and oxide and can be wet chemically etched in a \SI{40}{\%} \ce{KOH} solution at \SI{50}{\celsius} for \SI{60}{min} to obtain a pedestal height of approximately \SI{10}{\micro m}. For fabricating the mold structure on this pedestal, the e-beam lithography process explained before is repeated, including the reactive ion etching with \ce{CHF3} and \ce{O2} ions and resist stripping in \ce{O2} plasma. Here, a current of \SI{50}{nA} is used for the exposure to get a slightly higher resolution.

\begin{figure*}[t]
    \includegraphics{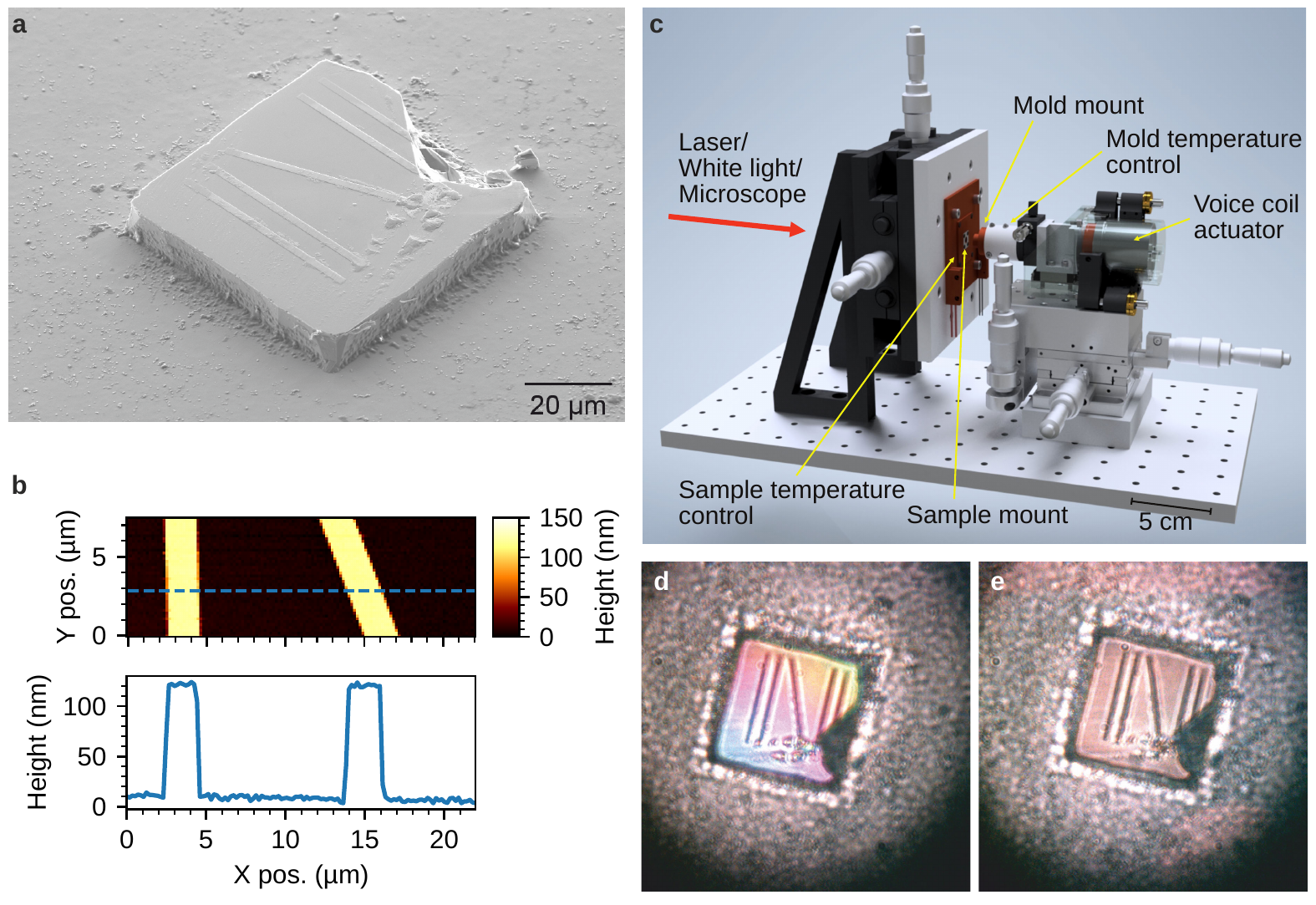}
    \caption{Scanning electron miroscope (SEM) image of a fabricated mold consisting of pedestal with bars on top. (b) AFM image and height profile of the central bars. (c) Illustration of the nanoimprint setup. (d) Optical reflection micrograph of a pedestal positioned slightly above the sample. Colors are due to optical interference of the light between the mold and monolayer surface and are used for parallel alignment. (e) Optical reflection micrograph of the mold in contact with the monolayer.}
    \label{fig:imprint_method}
\end{figure*}

Figure \ref{fig:imprint_method}a shows a scanning electron microscope image (SEM) of a typical mold consisting of five bars sitting on a 10 µm high pedestal with lateral dimensions of \SI{67}{\micro m} $\times$ \SI{67}{\micro m}. All bars have a width of \SI{2}{\micro m}. The height of the three central bars is \SI{110}{nm}, and the height of the two outer bars is \SI{80}{nm} due to a lower thickness of the resist at the outer part of the pedestal during the second spin coating step. The SEM image of the mold has been recorded after about 100 imprints, demonstrating the reusability of the mold, with only a triangular-shaped piece of the edge of the mold broken off.

The mold is used to imprint the pattern into the monolayer \ce{WS2}/PMMA/fused silica sample using a custom-built experimental setup displayed in Fig.~\ref{fig:imprint_method}c. It consists of a sample holder with a micrometer positioning stage and a sample heater to control the temperature of the sample during the imprint process. Alignment of the mold with respect to the target substrate is achieved by a kinematic mounting, which can be adjusted in x, y and z direction as well as in pitch and yaw. The mold can also be heated up to \SI{200}{\celsius}. To imprint the mold into the sample with a controlled and constant force, we use a voice coil actuator. On the back of the sample holder, an optical microscope (not shown) is installed to adjust the sample and mold position with respect to each other. For a successful imprint, a parallel alignment of the mold and the substrate is crucial. To ensure this, we use a paraxial illumination scheme in the microscope creating optical interference of the light between the pedestal surface and the sample. The interference can be used to adjust the tilt of the mold while monitoring the color gradients (Fig.~\ref{fig:imprint_method}d). In that way, we achieve a parallel alignment with an angular deviation below \SI{0.1}{\degree}.  

The imprint process is optimized by adjusting the force and temperature during the imprint. To perform an imprint at relatively low force, the sample is heated to \SI{180}{\celsius} to soften the PMMA layer, whose glass transition temperature is around \SI{105}{\celsius}. After this, the mold is pressed into the TMDC monolayer with a force of \SI{1.9}{N}. Subsequently, the heating of the substrate is switched off, and the mold pulled out after the substrate has cooled to a temperature of \SI{70}{\celsius}. 

\begin{figure*}[tb]
    \includegraphics{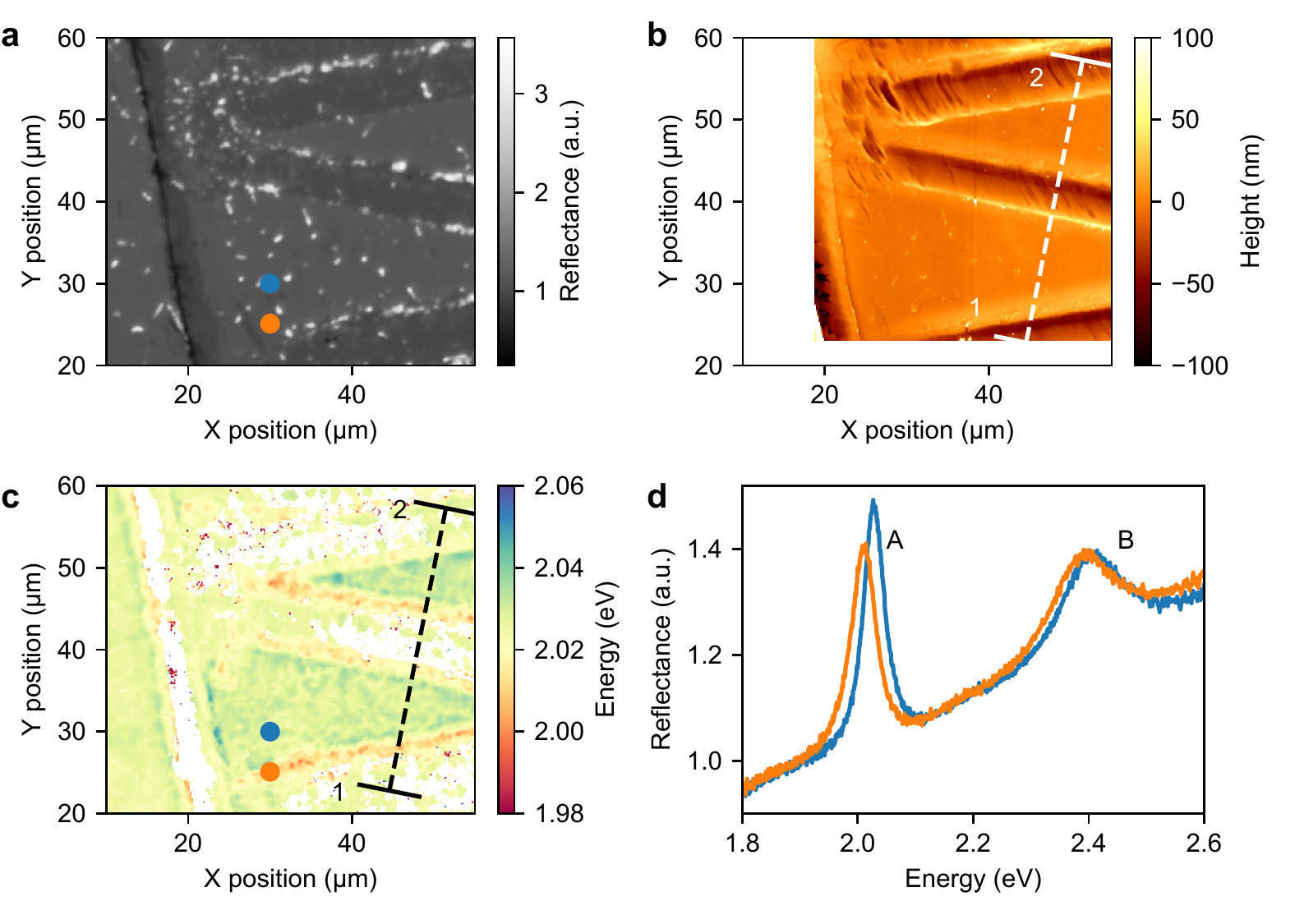}
    \caption{Creation of an inhomogeneous strain profile in a \ce{WS2} monolayer (a) Optical reflectance map of an imprinted area with three bars. Profiles along the dashed white line are shown in Figure \ref{fig:profiles}. (b) Atomic force microscope (AFM) image of the same region as in (a), showing the depth of the imprint. Zero height is set at the center between the indented regions (close to the blue dot in (a) and (b)). (c) Energy map of the A exciton. In the white areas, the \ce{WS2} monolayer is absent. (d) Reflectance spectra at the two locations marked with colored dots in (c).}
    \label{fig:results}
\end{figure*}

To measure the created strain distribution in the atomically thin TMDC material, we perform spatially resolved transmission and reflection spectroscopy (hyperspectral imaging \cite{Darlington2020Imaging}) of the fabricated samples in a custom-built microscopy setup (see supporting information and Fig.~\ref{fig:mapping_setup} for details). We extract the energy of the A exciton of the TMDC monolayer from the acquired spectra for each location, as a measure for the amount of induced strain \cite{Schmidt2016Reversible,Rosati2021Dark}. To map the strain in the 2D material, the sample is scanned, and at each position transmission and reflection spectra are recorded. Afterwards, the spectra are normalized to the transmission and reflection spectrum at a location where only the bare substrate with the PMMA layer is present, respectively.

A peak finding algorithm is used to extract the energy of the energetically lowest exciton (the A exciton) of every scanned pixel and create an exciton energy map of the sample. This map yields information on the amount of local strain present in the 2D material.

\begin{figure*}[!ht]
    \includegraphics{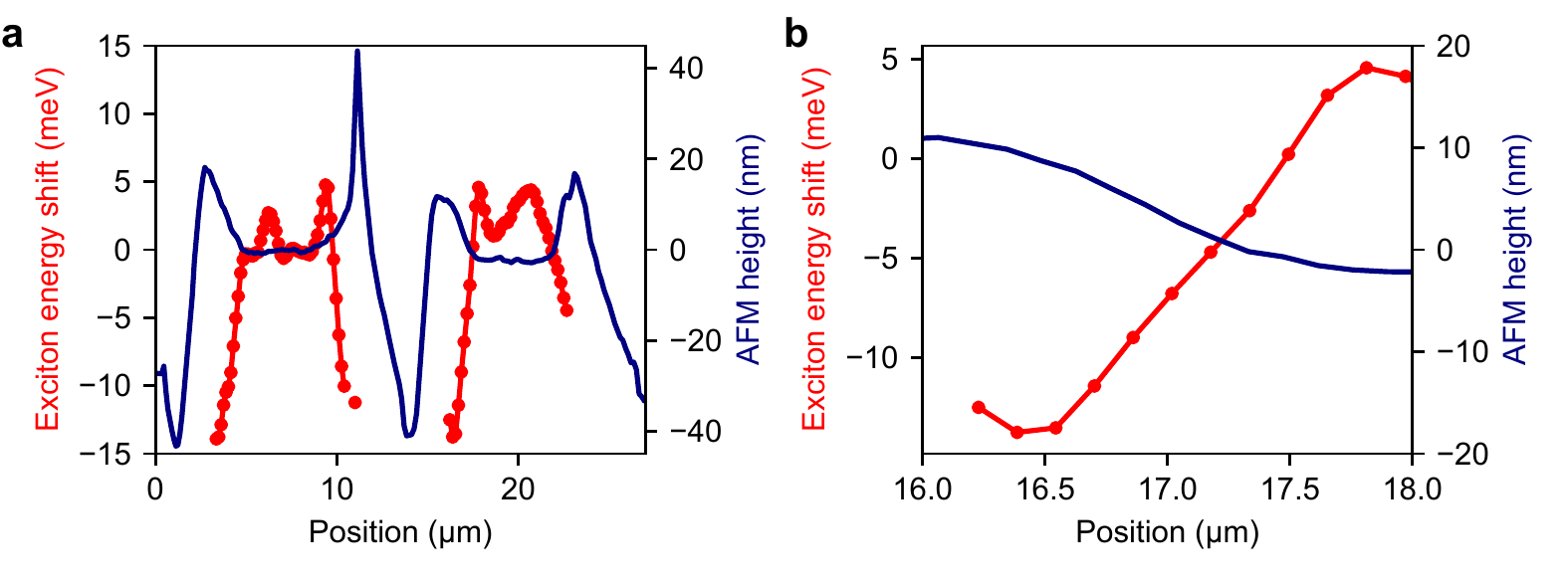}
    \caption{Strain gradients in the \ce{WS2} monolayer (a) Line cuts of the AFM height (blue) and the relative shift of the A exciton energy (red), along the dashed lines in Figure~\ref{fig:results}b and c. (b) shows a zoomed region of (a).}
    \label{fig:profiles}
\end{figure*}

Figure \ref{fig:results} shows the results of a typical imprint into a CVD-grown \ce{WS2} monolayer. The optical reflectance map of an imprinted region of the monolayer on the PMMA/fused silica substrate is displayed in Fig.~\ref{fig:results}a. Here, the dark grey, groove-like regions are imprinted, while the lighter regions are due to the pristine monolayer. White dots are either nucleation centers from the CVD growth or parts of the \ce{WS2} monolayer, which are folded upwards \cite{Senthilkumar2014Direct}. The reflectance map (Figure \ref{fig:results}a) contains information about the topography as well the optical response of excitonic resonances. To disentangle these contributions, we record an atomic force microscopy (AFM) image of the imprinted region (Figure \ref{fig:results}b). At the position of the imprint, the monolayer was pressed into the substrate by \SI{40}{nm} (cf. Fig.~\ref{fig:profiles}), which is \SI{36}{\%} of the height of the stamp structures. Since the mold is in contact with the PMMA layer during the imprint process, we attribute this to the elastic response of the polymer, when the mold is removed. Next to the imprinted regions, the AFM height map shows elevated, wall-like regions, which are built up by polymer material redistributed from the groove. Furthermore, scratch-like parallel structures are visible in the indented regions. These are caused by a lateral movement between the sample and the mold during the cooling phase of the imprint, originating from the thermal expansion of the substrate. 

The imprint causes strain in the \ce{WS2} monolayer. Mechanical strain results in a shift of the A exciton resonance with a gauge factor of \SI{-50}{meV/\%} for uniaxial strain \cite{Niehues2018Strain}.  A lower exciton energy corresponds to larger tensile strain. The A exciton energy, extracted from the hyperspectral map, is displayed in Figure \ref{fig:results}c. In the image, orange to red colors represent red-shifted A exciton energies. The white areas apparent in some regions of the imprinted grooves and the edge of the mold indicate that no exciton can be detected due to the destruction or removal of the \ce{WS2} monolayer by the mold. We find that in regions close to the imprint the exciton energy is red-shifted in an area with a width of $\approx \SI{2}{\micro m}$ with respect to the plateau between the imprinted grooves. For illustration, in Figure \ref{fig:results}d the reflectance spectra of the monolayer at the locations indicated by the blue and orange dots in Figs. \ref{fig:results}a and c are shown. Comparing the orange to the blue curve, a 17 meV red-shift of the A exciton resonance is visible, which corresponds to a uniaxial tensile strain of \SI{0.34}{\%}. As expected, the B exciton also shifts towards lower energies.

To explain the created strain, we extract the energy shift of the A exciton from Fig. \ref{fig:results}c and the topography profile from Fig.~\ref{fig:results}b, both along the dashed lines in Fig. \ref{fig:results}b and c, and plot them in Fig.~\ref{fig:profiles}a. The AFM height profile in Fig. \ref{fig:profiles}a starts in the lower imprinted area (1), crosses the first plateau, the second (center) imprint, and the second plateau. The endpoint (2) of the profile is located in the third imprint. In order to increase the signal-to-noise ratio, the profiles are averaged over an area indicated by the cap length of the profile lines in Figs. \ref{fig:results}b and c. Negative values in the AFM profile are regions where the mold was pressed into the monolayer/polymer, while positive values denote areas, where monolayer/polymer material was redeposited during the imprint, creating dike-like structures. The asymmetry of the imprint profiles results from a slight drift of the sample with respect to the mold during the temperature ramping of the imprint process. By comparing the AFM height profile and the energy shifts of the A exciton of Figure \ref{fig:profiles}a, it is evident, that the strong energy shifts ($> \SI{5}{meV}$) appear in regions where the polymer has been re-deposited. The almost symmetric, negative exciton shift on both sides of the central imprinted structure at \SI{10}{\micro m} and \SI{17}{\micro m}  indicates, that the material movement during the imprint and not the drift of the sample while imprinting is the reason for the strained monolayer. Fig. \ref{fig:profiles}b shows a zoomed-in region of Fig. \ref{fig:profiles}a close to the imprint. The imprint has created an almost linear gradient of the exciton energy over a distance of \SI{1.5}{\micro m} with a slope of \SI{15}{meV/\micro m}. Assuming uniaxial strain in the \ce{WS2} monolayer, this value corresponds to a strain gradient of \SI{0.3}{\%/\micro m}.

In conclusion, we have demonstrated that an in-situ imprint technique can be used to create strain gradients in the micrometer range in a \ce{WS2} monolayer. Additonal control over the amount of strain and the shape of the gradients can be envisioned by optimizing imprint parameters or designing different 3D patterns on the mold. Our technique is scalable and could be transferred to commercial nanoimprint machines. This is an important step towards the future application of strained 2D materials in commercial optoelectronic devices.

\begin{acknowledgements}
We gratefully acknowledge funding by the German Science Foundation (DFG) via project number 231996899. We thank the Münster Nanofabrication Facility (MNF) for their support in nanofabrication-related matters. We thank Helge Gehring, Corinna Kaspar, and Wolfram Pernice for their help in an early stage of this study. 
\end{acknowledgements}

%\nocite{*}

\bibliography{ref}% Produces the bibliography via BibTeX.

%%%%%%%%%% Merge with supplemental materials %%%%%%%%%%
\clearpage
\widetext
\begin{center}
\textbf{\large Supplemental Material}
\end{center}
%%%%%%%%%% Merge with supplemental materials %%%%%%%%%%
%%%%%%%%%% Prefix a "S" to all equations, figures, tables and reset the counter %%%%%%%%%%
\setcounter{equation}{0}
\setcounter{figure}{0}
\setcounter{table}{0}
\setcounter{page}{1}
\makeatletter
\renewcommand{\theequation}{S\arabic{equation}}
\renewcommand{\thefigure}{S\arabic{figure}}
\renewcommand{\bibnumfmt}[1]{[S#1]}
\renewcommand{\citenumfont}[1]{S#1}
%%%%%%%%%% Prefix a "S" to all equations, figures, tables and reset the counter %%%%%%%%%%

\section*{Mold fabrication process}

\begin{figure}[!h]
    \centering
    \includegraphics{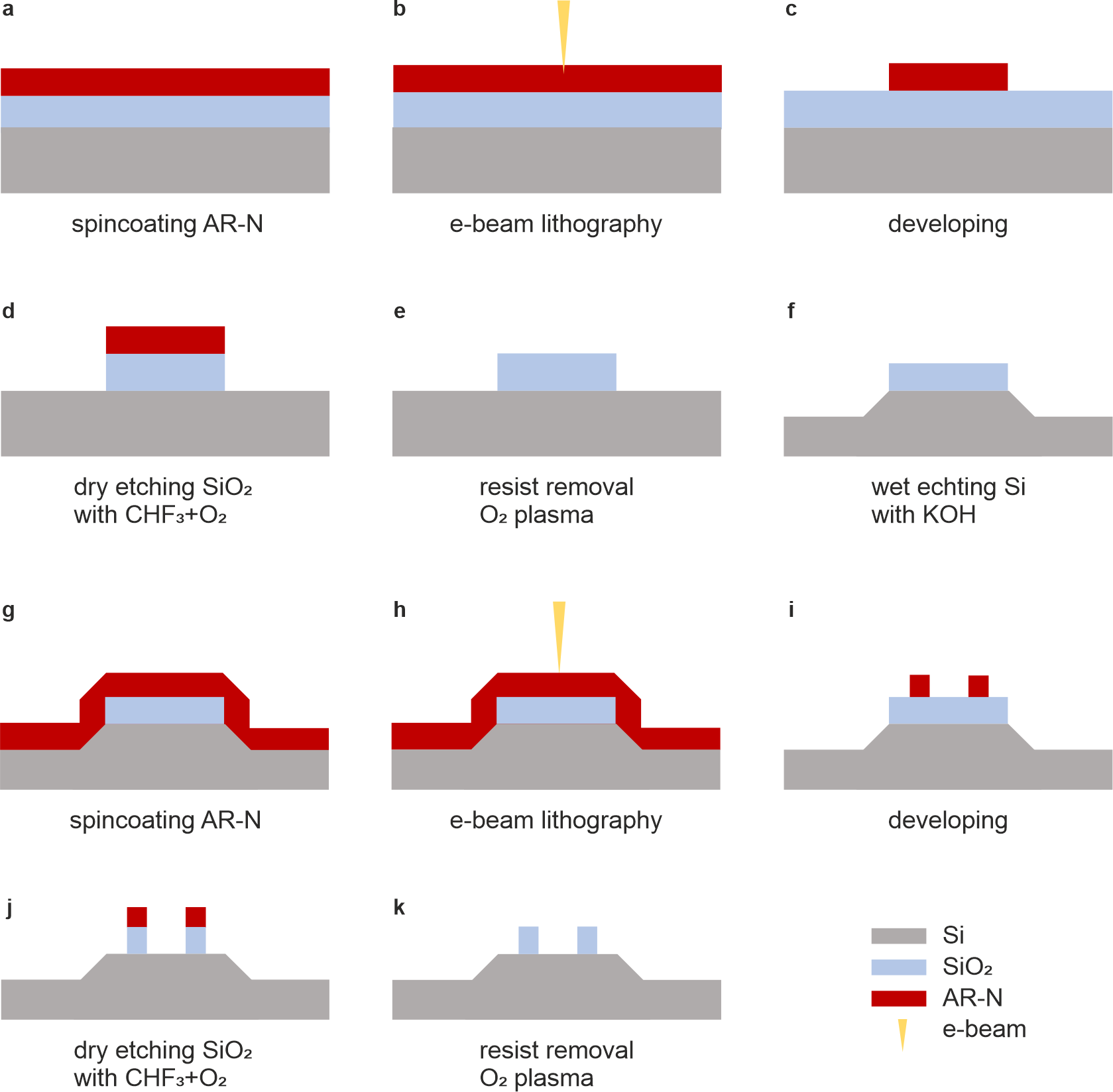}
    \caption{Mold fabrication process}
    \label{fig:process}
\end{figure}

\section*{Hyperspectral optical imaging setup}

\begin{figure*}[!ht]
    \includegraphics{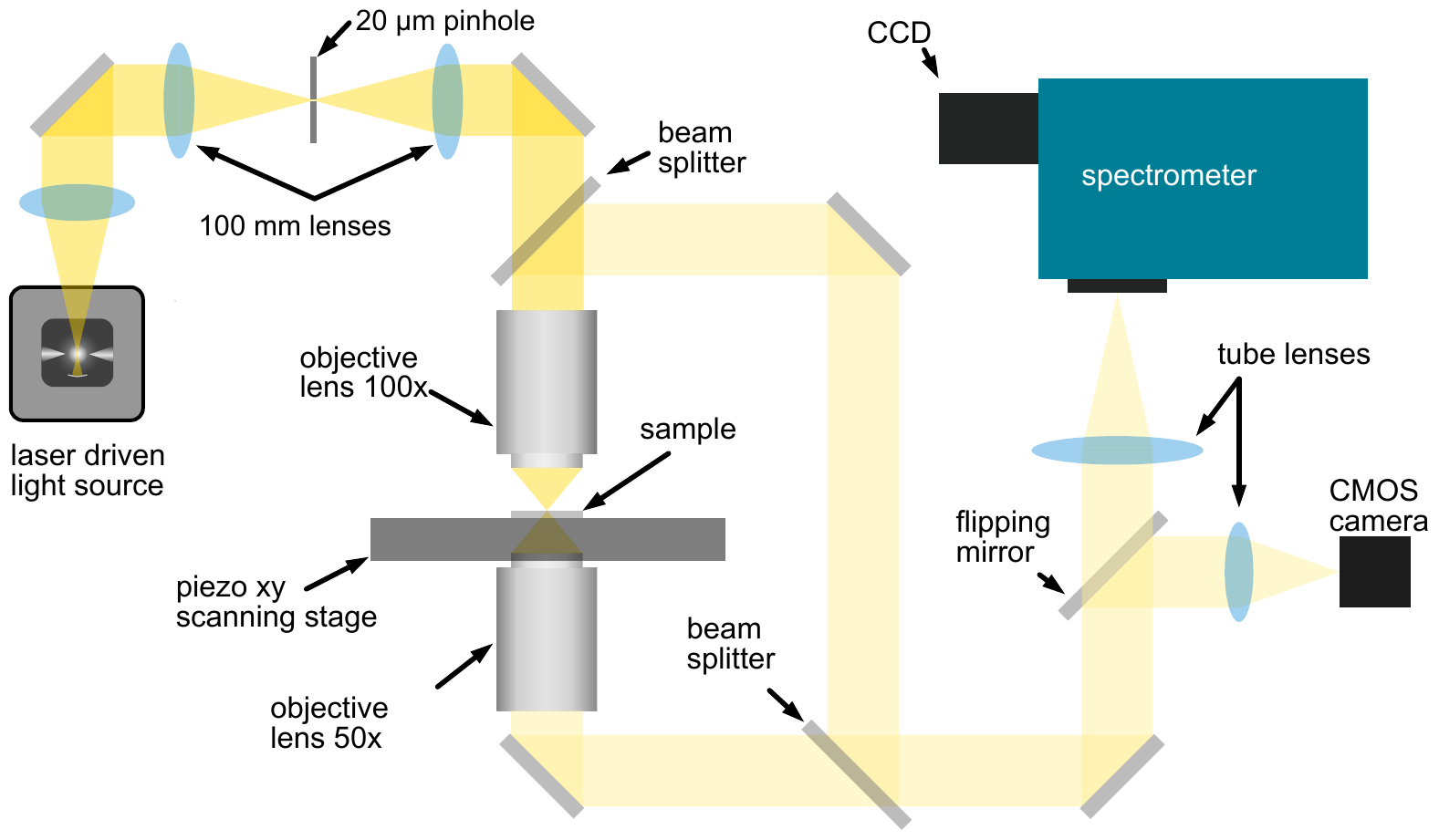}
    \caption{Schematic drawing of the hyperspectral optical imaging setup to map the strain in the imprinted 2D materials.}
    \label{fig:mapping_setup}
\end{figure*}

Figure \ref{fig:mapping_setup} shows a schematic drawing of the hyperspectral imaging setup. Light from a high-power laser driven light source (LDLS, Energetiq EQ-99X) is collimated using an achromatic lens (\SI{50}{mm}). The collimated light is focused onto a \SI{20}{\micro m} pinhole and recollimated with two identical \SI{100}{mm} lenses. Subsequently, it is focused onto the sample using an objective lens (Nikon TU Plan 100x, $NA=0.9$), creating a nominal spot size of \SI{400}{nm}, which defines the resolution of the setup. The response from the sample is collected simultaneously in reflection and transmission geometry by two objective lenses (reflection: Nikon 100x Plan Flour, numerical aperture $NA = 0.9$; transmission: Nikon 50x TU Plan Epi ELWD, $NA=0.7$). The sample is precisely scanned using a piezo xy-scanning stage (MadCityLabs NanoT22) with a maximum travel range of \SI{200}{\micro m} in X and Y direction. A CMOS camera is used for orientation on the sample. For strain mapping, the reflected and transmitted light are imaged onto the entrance slit of a spectrometer (Andor Shamrock SR-i303) equipped with a charge-coupled device (CCD, Andor Newton DU920P-OE) using an achromatic doublet lens with a focal length of \SI{400}{mm}.

\end{document}